\documentclass[a4paper,11pt]{article}

\usepackage{jheppub} 

\usepackage[T1]{fontenc} 
\usepackage{graphicx} 
\usepackage{bm}
\usepackage[mathlines]{lineno}
\usepackage{amsmath} 
\usepackage{hyperref}
\usepackage{amssymb}
\usepackage{verbatim}
\usepackage{amstext} 
\usepackage{amsfonts}
\usepackage{braket}
\usepackage{MnSymbol}

\newcommand{\intd}{\int \text{d}}

\begin{document}
	\title{\boldmath On the domain of moduli fields}
	
	\author[1]{Thomas Steingasser}
	
	
	\affiliation[1]{Ludwig-Maximilians-Universit\"at,\\Theresienstra\ss e 37, Munich, Germany}
	
	\emailAdd{Thomas.Steingasser@physik.lmu.de}

	\abstract{The concept of the moduli space allows for a simple, universally applicable description of the low-energy dynamics of topological solitons. This description is remarkably insensitive to the properties of the underlying theory, whose details only manifest themselves via the moduli space metric. This article presents a generalization of this concept, which allows to transfer its most intriguing features to configurations of any energy captured by the theory giving rise to the soliton, given that these are localized sufficiently close to the soliton's center. The resulting theory is capable of describing all dynamics within its range of applicability by just one family of fields, with all the information about the underlying theory entering via a finite number of background functions, which can be linked to physical properties of the present soliton. }

	\maketitle
	\flushbottom
	\newpage
	\section{Introduction}
	This article sets out from the connection between three important aspects of modern theoretical physics: Topological solitons and the theory of fluctuations around them, Goldstone's theorem and the philosophy of effective field theory. \newline
	The semi-classical quantization of topological solitons has established itself as a standard procedure, not only due to its clear, intuitive comprehensibility, but also because of its great successes in the description of real physical systems. Topological solitons can be found in many fields of modern physics, from the effective description of baryons over string theory to solid state and condensed matter physics as well as in cosmology \cite{Proton}-\cite{Brane}.  \newline
	As all solitons break symmetries of their underlying theories, the spectra of fluctuations around them contain non-trivial zero modes associated to these broken symmetries. Over the last decades, several approaches to handle these modes have been developed \cite{creutz1975}-\cite{jackiw1977}. While physical results do not depend on the chosen approach, each of them is associated with an own perspective and has the potential to simplify certain computations. 
	\newline
	One way to understand these modes is Goldstone's theorem. This theorem, which links spontaneously broken symmetries to massless excitations, is one of the most consequential and universal results of the second half of the $\text{20}^{\text{th}}$ century. Its applications reach from the foundations of pion physics to solid state physics, shaping the low energy dynamics of the systems it is applicable to. \newline
	These Goldstone modes dominate the low-energy behavior of the theories they arise from. They therefore play a central role in the effective theories describing the theories from which they arise at energies lower than the masses of these theories' excitations. One of the main motivations behind the investigation and construction of such effective field theories is that they often provide a drastically simpler picture of the systems they are used to describe, thereby pointing at otherwise hidden aspects of their underlying theories. A particularly important example for such a setup in the context of topological solitons is the Skyrme model, which can be understood as an effective theory of QCD in terms of its Goldstone modes in the form of pions. In this theory, baryons manifest themselves as topological solitons, so-called Skyrmions, and many of their most important properties can be obtained already on the classical level \cite{Proton}.
	
	\subsection{Motivation \& Overview}
	Moduli fields play an essential role in all discussions of solitons. Their most appealing property is thereby that their dynamics is independent of most of the details of the underlying theory, including but not restricted to the precise shape of the soliton, the types of fields involved in its formation, the interactions between them and even the theory's topological properties. On this level, the information about all these aspects is fully contained in the moduli space metric. In other words, knowing the number of moduli, all that is necessary to obtain a full description of the system's low energy dynamics is a small number of real parameters.  \newline
	This universality breaks down as soon as non-zero modes become involved, as the theory of the non-zero modes strongly depends on the type of involved fields as well as on their interactions. This article presents an approach capable of preserving the universality of the description in terms of moduli fields, while also including a subsector of the dynamics associated with the non-zero modes. This approach is valid for all energies allowed by the underlying theory, as long as the considered configurations are localized sufficiently close around the soliton's center.  \newline 
	The presented procedure is based on the introduction of so-called \textit{warp fields}, which can be obtained by promoting a sufficient number of the theory's moduli fields to functions on all of spacetime. Just as the moduli fields, their main feature lies in their universality, as the general structure of their theory is almost entirely insensitive to any peculiarities of the underlying theory. All the information regarding these enters only via a finite number of background functions, taking the role of a generalized moduli space metric. All of this is in particular valid for the full, non-linear theory.\newline
	The crucial step performed in this article is the extension of moduli fields to functions on all of spacetime. In the context of collective coordinates, this can be broken down to the promotion of collective coordinates to fields. As this is an unusual approach, this article sets out from a simple example, demonstrating the physical meaning of this process, afterwards carefully and rigorously filling this notion with meaning. To avoid the complications of higher-dimensional models and the presence of gauge fields, all of this is first done in full detail for the case of a (1 + 1)-dimensional system. \newline 
	Sections 2 provides an extensive discussion of the properties of the theory obtained in the way described above. As the warp fields are obtained from the theory's Goldstone modes, the essential features characteristic for such fields can be generalized such that they are also shared by the warp fields. One of these properties is the localization around the soliton's center, which is discussed in sections 2.4 and 2.5. \newline 
	The most remarkable of these properties is highlighted in section 2.7 for the $(1+1)$-dimensional case. It is well-known that gauging the symmetry whose breaking induces a Goldstone field leads to the latter being eaten up by the gauge boson corresponding to the symmetry. Due to the warp field's connection with the translational Goldstone field, it can be absorbed into the spacetime's metric, a procedure within which the warp field is replaced as a degree of freedom by a scalar field appearing during the construction.  \newline 
The idea of warp fields is not limited to classical field theories, but can also be lifted to the quantum level. For completeness, the quantum theory of the warp field in $(1+1)$ dimensions is presented in the appendix. \newline
The $(1+1)$-dimensional case is however far from generic, as it naturally avoids the greatest difficulty of this approach. Its major restraint results from the pivotal connection between the warp fields and the system's spatial dimensions. The construction given in section 2 can only be applied without additional efforts if the number of the system's real degrees of freedom is equal to the number of translational symmetries broken by the soliton. Section 3.2 is dedicated to the construction of the Lagrangian of any such theory in terms of warp fields starting from the moduli space, and presents relations between the appearing background functions and physical properties of the underlying theory. This general discussion is illustrated on the example of the Skyrme model in $3+1$ dimensions in section 3.3, which is discussed in great detail. \newline 
	The Skyrme model serves as a suitable example for this setting as the number of its independent, real degrees of freedom happens to match the number of spatial dimensions. It is further particularly interesting as it contains terms of higher orders in derivatives of its fields, allowing to showcase a common generalization of the results obtained in the previous subsection.  \newline
	For a generic theory however, this matching does not necessarily have to hold. It is argued in Section 3.4 that supplementing the warp fields with additional fields constructed from additional zero modes allows to extend the previous constructions to a much wider range of theories. One such example is the abelian Higgs model in $2+1$ dimensions, for which such a construction is given in full detail in section 3.5.. The solitons of this system break two translational symmetries, while containing three degrees of freedom. This mismatch can however be accounted for by including the system's third massless mode, which is linked to the system's $U(1)$-symmetry. 
	
	\subsection{Incitement: Local interactions}
	To get a better intuition regarding the physical meaning of the warp field, consider the example of a $(1+1)$-dimensional soliton together with its collective coordinate, $\Phi_s (x-z(t))$. While the collective coordinate is only a function of time, the corresponding fluctuation extends over all of spacetime, decaying as $\Phi^\prime_s $. It is therefore predominantly localized in the same spatial region as the soliton itself. Assume further that this soliton is interacting with some external, localized source, consisting of zero as well as non-zero mode contributions. Such an interaction changes the equation of motion of $z(t)$, therefore changing $\dot{z}(t)$. As $z(t)$ is only a function of time, but the fluctuation induced by it extends over all of space, the information about the excitation seems to spread instantaneously, violating causality. \newline
	However, there is in fact no problem here. For the external source to be localized, it has to be a superposition of the zero as well as the non-zero modes, so that such an interaction would excite all the involved modes. The resulting excitation can therefore without any problems be localized within the region of the interaction. Afterwards, it can be expected that the excitation spreads out with some finite velocity smaller than 1. This would imply that immediately after the interaction begins, the soliton only moves in the neighborhood in which the interaction is taking place. This excitation then starts spreading with some finite velocity, causing larger and larger parts of the soliton to move. As different regions of the soliton are moving with different velocities, the soliton is warped during this process. This can be formalized by replacing the collective coordinate by a local object,
	\begin{align}
	z(t) \to \mu_m^{-1} \varphi(t,x),
	\end{align}{}
	which is defined with an additional factor of $\mu_m$ to obtain a dimensionless quantity. This object will from now on be referred to as \textit{warp field}. \newline
	The velocity of the motion in $x$-direction \textit{in some point} $x_0$ is then given by $\mu_m^{-1} \dot{\varphi}(t,x_0)$. This velocity can be expected to consist of two contributions: First, a \textit{global} velocity $v$, which is the result of the dynamics of the collective coordinate. Second, a \textit{local} velocity, which is fully determined by the dynamics associated with the non-zero modes. \newline
	An alternative perspective on such a configuration is based on the soliton's width. It is well known that providing any soliton-like configuration with a sufficient amount of energy will cause a (space)-time dependent perturbation of its thickness \cite{Shifman}, which will eventually decay into separate particles. The warp fields can be understood as an attempt to describe this phenomenon in a compact manner.
	
	\subsection{Notation and Conventions}
	For simplicity, let the potential generating the soliton in the $(1+1)$-dimensional case be symmetric, $V(\Phi)=V(-\Phi)$, with two distinct vacua $ \pm \nu$. Further, assume that the soliton carries a topological charge 1, for which it is always possible to find a coordinate frame in which $\Phi_s = \Phi_s (x)$. This allows for the following representation:
	\begin{equation}
	\Phi_s(x) = \nu \sigma (\mu_m x )
	\end{equation}
	Here, $\nu$ corresponds to the dimensionless vacua of the theory as mentioned above, and the factor $\mu_m \propto m_\Phi$ describes the localization of the soliton. $\sigma$ encodes the profile of the soliton, and is in general restricted by $|\sigma (\mu_m x)|<1$. $\sigma^\prime$ denotes the derivative of $\sigma$ with respect to its dimensionless argument. The same is true for partial derivatives acting on solitons in higher dimensions. A general fluctuation of this background will be denoted by $\delta \Phi$, $\Phi (t,x) = \Phi_s (x) + \delta \Phi (t,x)$. A fluctuation generated by non-zero modes only will be denoted by $\phi$, so that $\Phi (t,x) = \Phi_s (x -z(t)) + \phi (t,x)$. \newline
	Throughout this article, two different perspectives are explored: One, in which the dynamical degrees of freedom are given by $\phi$ and the collective coordinate, and one, in which this role is taken by the warp field $\varphi$. In the first case, the classical dynamics takes place on the phase space of $\phi$, denoted by $\mathbb{P}$. The quantum theory unfolds on the Fock space the field operators act on, $\mathbb{F}$. When using $\varphi$ as degree of freedom, the corresponding phase/Fock space is denoted $\mathcal{P}/\mathcal{F}$. \newline
	The linearized equation of motion for $\phi$ is in general of the form
	\begin{equation}
	    \big(\partial_t^2 + K^2\big) \phi = 0,
	\end{equation}{}
where the operator $K^2$ contains only derivatives with respect to the position $x$ as well as functions of $x$. The eigenfunctions of this operator will be denoted by $\{ f_k \}_k$, so that $\phi$ can be expanded as $\phi = \sumint_k a_k^\dagger e^{i \omega_k t} f_k^* + c.c. $, with some complex coefficients $\{ a_k \}_k$. The symbol $\sumint$ is meant as sum over the discrete part of the spectrum and integration over the continuous part, if need be containing some normalization factors depending on the normalization of the eigenfunctions or momentum eigenstates in the quantum theory.

\section{Classical theory of the warp field in 1+1 dimensions}

\subsection{Foundations of the classical theory}
The action of the warp field can be obtained by inserting the warped soliton into the action of the scalar field. Doing so, one finds 
\begin{equation}
	\begin{split}
	S=  \intd ^2x & \ \frac{1}{2} \nu^2 \big(\sigma^\prime (\mu_m x - \varphi) \big)^2 \big( \dot{\varphi}^2 - (\varphi^\prime)^2-\mu_m^2 + 2 \mu_m \varphi^\prime \big) - \\ 
	&-V\big(\nu \sigma (\mu_m x -\varphi) \big). \label{Action0}
	\end{split}{}
\end{equation}{}
Introducing the dimensionless coordinates $\xi = \mu_m x - \varphi$, Bogomolnyi's equation implies that $\sqrt{2 V(\xi)}= \mu_m \nu \sigma^\prime (\xi)  $. Thus, the potential can be eliminated from \eqref{Action0}, which can then be brought to the form
	\begin{equation}
	\begin{split}
	S= \intd ^2x & \frac{1}{2} \nu^2 \big(\sigma^\prime (\xi) \big)^2  \partial_\mu \varphi \partial^\mu \varphi + \\ 
	&+ \nu^2 \big(\sigma^\prime (\xi) \big)^2 \big(-\mu_m^2 + \mu_m \varphi^\prime \big) . \label{Action2}
	\end{split}{}
	\end{equation}{}
	Consider now the terms in the second line. The spatial integral can be rewritten as
	\begin{equation}
	\begin{split}
	\intd x \ & \nu^2 \big(\sigma^\prime (\xi ) \big)^2 \big(-\mu_m^2 + \mu_m \varphi^\prime \big) = - \mu_m \intd x  \frac{d\xi}{dx} \  \nu^2 \big(\sigma^\prime (\xi) \big)^2  = \\ 
	=& -  \intd \xi \ \mu_m \nu^2 \big(\sigma^\prime (\xi) \big)^2 = -  M_{\text{sol}} = L[\Phi_s]
	\label{Action3}
	\end{split}{}
	\end{equation}{}
	
	Thus, the action of the warped soliton can be brought to the simple form
	\begin{equation}
	\begin{split}
	S= S[\Phi_s]+ \intd ^2x & \frac{1}{2} \nu^2 \big(\sigma^\prime (\xi) \big)^2 \partial_\mu \varphi  \partial^\mu \varphi , \label{Actionf}
	\end{split}{}
	\end{equation}{}
	which is valid in full non-linearity. \newline 
	This action can be simplified even further by absorbing the soliton density into the spatial measure\footnote{Note that while this observation is not necessary for the understanding of the classical theory, it is crucial for the construction of the quantum theory.}, so that
	\begin{equation}{}
	\begin{split}
	S [\varphi] =& \frac{1}{2} \intd ^2\varsigma_\varphi (x)   \partial_\mu \varphi  \partial^\mu \varphi,\ \text{where} \\
	\text{d}^2\varsigma_\varphi (x) =&\nu^2 \big(\sigma^\prime (\mu_m x - \varphi) \big)^2 \text{d}x \text{d}t .
	\end{split}
	\end{equation}
	Therefore, the action takes the same form as the one of a usual massless scalar field, weighted with the function $\rho (x, \varphi (t,x)) = \nu^2 \big(\sigma^\prime (\xi) \big)^2$. In other words, this approach allows - within the boundaries of its applicability, which will be specified in section 2.4 - to treat all the theory's interactions by a non-trivial replacement of the spatial integration measure. \newline
	In order to allow for a standard perturbative treatment of this theory, one would have to expand the term $\big(\sigma^\prime (\mu_m x - \varphi)\big)^2$ as a series in $\varphi$. This leads to a kinetic term as well as an infinite number of interaction terms,
	\begin{equation}{}
	\begin{split}
	\mathcal{L}_0=&\frac{1}{2} \nu^2 \big(\sigma^\prime (\mu_m x) \big)^2 \partial_\mu \varphi  \partial^\mu \varphi, \\
	\mathcal{L}_{\text{int}}^{(n)}=&\frac{(-1)^n}{2 \cdot n!} \nu^2 \big( \big(\sigma^\prime (\mu_m x ) \big)^2\big)^{(n)}   \varphi^n \partial_\mu \varphi  \partial^\mu \varphi.
	\end{split}
	\end{equation}{}
	Here, the power ${(n)}$ refers to the $n^{\text{th}}$ derivative with respect to the dimensionless argument $\mu_m x$. Note that, similar as for a Goldstone field, each of these terms disappears in the limit $\partial \varphi \to 0$, which is to be expected due to the warp fields relation to the collective coordinate.  \newline
	This Lagrangian gives rise to the equation of motion,
	\begin{equation}
	\Box \varphi = \dfrac {\sigma^{\prime \prime}(\xi)}{\sigma^\prime(\xi)}  \big(  \partial_\mu \varphi \partial^\mu \varphi + 2 \mu_m  \varphi^\prime \big),
	\label{EOMWarp}
	\end{equation}
	and determines the canonical momentum $\varpi$,
	\begin{align}
	\varpi (t,x) = \nu^2 \big(\sigma^\prime (\xi) \big)^2 \dot{\varphi} (t,x)=\rho (x, \varphi (t,x)) \dot{\varphi} (t,x).
	\end{align}{}
	The phase space on which the dynamics of the warp field takes place, spanned by $\varphi$ and $\varpi$, will in the following be denoted by $\mathcal{P}$. \newline
	Another important feature of the theory is its energy-momentum tensor, which can be found to be 
	\begin{equation}{}
	\begin{split}
	T^\mu_\nu=& \nu^2 \big(\sigma^\prime (\xi)\big)^2 \bigg(\partial^\mu \varphi \partial_\nu \varphi - \delta^\mu_\nu \frac{1}{2}  \partial^\alpha \varphi \partial_\alpha \varphi \bigg)=\\
	=&\rho (x, \varphi (t,x))\bigg(\partial^\mu \varphi \partial_\nu \varphi - \delta^\mu_\nu \frac{1}{2}  \partial^\alpha \varphi \partial_\alpha \varphi \bigg).
	\end{split}
	\end{equation}{}
	Note that this expression is again just the usual expression for a massless scalar field multiplied by the weight factor.

	\subsection{Linearized theory}
	
	In order to avoid the technical difficulties arising from the infinite number of interaction terms and prepare for the canonical quantization of this system, it is expedient to focus on the linearized theory. This is a good approximation if $\varphi$ satisfies the inequality
	\begin{align}
	|\varphi (t,x) | \ll \bigg| n! \frac{\big(\sigma^\prime  (\mu_m x ) \big)^2}{\big((\sigma^\prime (\mu_m x ))^2 \big)^{(n)} } \bigg|^{1/n} \ \forall n\in \mathbb{N} \label{approx},
	\end{align}
	which can be obtained from the above action. In the case of the $\Phi^4$-theory, the right hand side of this equation is bounded from below by $\frac{1}{4}$, in the Sine-Gordon model by $\frac{1}{2}$. \newline
	In this approximation the action and equation of motion reduce to
	\begin{equation}
	\begin{split}
	S[\varphi]  =& \intd ^2x \  \dfrac{1}{2 } \nu^2 \big(\sigma^\prime  (\mu_m x ) \big)^2  \partial_\mu \varphi \partial^\mu \varphi ,  \\
	&\big(\partial_t^2 + \mathcal{K}^2\big) \varphi = 0 \label{LinEOM}, \ \text{where} \\
	\mathcal{K}^2:=&-\partial_x^2 - 2 \mu_m \dfrac {\sigma^{\prime \prime}}{\sigma^\prime} (\mu_m x ) \partial_x.
	\end{split}{}
	\end{equation}
	The first step towards solving this equation is to investigate the operator $\mathcal{K}^2$. While it isn't hermitian with respect to the measure d$x$, it is with respect to the due to the linearization $\varphi$-independent measure d$\varsigma := \nu^2 \big(\sigma^\prime (\mu_m x )\big)^2 $d$ x$. Introducing the notation $\rho(x) \equiv \rho (x,0)$, this can be brought to the compact form d$\varsigma := \rho (x) $d$ x$. Hence, the eigenfunctions of $ \mathcal{K}^2$ form a complete basis, which can be orthonormalized with respect to d$ \varsigma$. 
	These eigenfunctions will be denoted by $\{g_k \}_k$, the corresponding eigenvalues by $\lambda_k$. It follows by a straightforward computation that these eigenfunctions are given by $g_k (x) = \frac{f_k(x)}{\nu \sigma^\prime (\mu_m x)} = \rho^{-1/2}(x) \cdot f_k (x)$, with $\lambda_k = \omega_k^2$. The functions $\{f_k\}_k$ denote here the eigenfunctions of the analogue of the operator $\mathcal{K}^2$ appearing in the $\phi$-theory, as defined in section 1.3. In terms of these eigenfunctions a general solution of \eqref{LinEOM} is of the form
	\begin{equation}
	\begin{split}
	\varphi (t,x)=   \sumint_{k \neq 0} & \alpha_k^* e^{i \omega_k t} g_k^* (x) + \alpha_k e^{-i \omega_k t} g_k (x) +  \mu_m v (t -t_0 ) + \mu_m \zeta, \label{Solvp}
	\end{split}{}
	\end{equation}
	with some real integration constants $v$ and $\zeta$ of mass dimension $0$ and $-1$ respectively. The sum/integral runs over all the non-zero modes, while the constant $\zeta$ represents the zero mode. The term linear in $ t $ is part of the kernel of $\partial_t^2$ as well as of $\mathcal{K}^2$, and describes a motion of the soliton with constant velocity. As $\varphi$ isn't canonically normalized, the additional factor of $\rho (x)= \nu^2(\sigma^\prime)^2 (\mu_m x)$ causes the contribution arising from these last two terms to yield a finite contribution to the action as well as all observables. \newline
	The expansion \eqref{Solvp} comes with an important subtlety. As the mass of the soliton is finite, $\sigma^\prime (\mu_m x) \to 0$ for $|x| \to \infty$, so that also $g_k \to \infty$ in this limit. It is important to notice that this does not necessarily imply that the same holds for $\varphi$, as it is possible to choose the coefficients $\{ \alpha_k \}$ such that $\varphi \to 0$ for $|x|\to \infty$. This observation is a necessary condition for the linearized theory to apply, as otherwise the condition \eqref{approx} would be violated. \newline
	This unusual asymptotic behavior of the mode functions will turn out to be of crucial importance for the quantum theory. The restriction \eqref{approx} also affects the term linear in $t$. Denoting the lower bound of its right hand side by $b$, it can be shown that the contribution of this term is agreement with \eqref{approx} if
	\begin{align}
	|v| \ll \frac{b}{\mu_m |T|},
	\end{align}{}
	where $T$ denotes the considered time interval, which is assumed to be centered around $t_0$. Higher orders can be obtained from standard perturbation theory, using $v$ as perturbation parameter. Starting from $\varphi^{(1)}= \mu_m v t$, the resulting perturbative series converges to 
	\begin{align}
	\varphi (t,x) = \mu_m \big( x - \gamma ( x-vt) \big), 
	\end{align}{}
	which reproduces the motion of the soliton for relativistic velocities,
	\begin{align}
	\Phi_s(x- \varphi (t,x) )= \Phi_s (\gamma (x-vt)). \label{HighO}
	\end{align}{}
	
	\subsection{Embedding into the full linearized theory}
	It remains to discuss which parts of the full theory's dynamics can actually be described in terms of the warp field. As the zero mode of $\varphi$ can be identified with the collective coordinate, this section will only focus on the non-zero modes. On the linearized level, introducing the warp field corresponds to the simple field redefinition
	\begin{align}
	\phi(t,x) \to - \nu \sigma^\prime (\mu_m x) \varphi(t,x). \label{Redef}
	\end{align}{}
	In order to relate expressions of the $\varphi$-theory to ones of the $\phi$-theory, it is expedient to extend this relation to a map between the phase spaces of the two theories. This map shall be denoted by $r$, and is given by
	
	\begin{equation}
	\begin{split}
	r: \mathcal{P} & \rightarrow \mathbb{P} \\ 
	\big( \varphi (x), \varpi (x) \big) & \mapsto \big(  r_\phi[\varphi ] (x) , r_\pi [\varpi ] (x) \big) \equiv \\
	&\equiv \big( - \rho^{1/2}(x) \cdot \varphi (x) ,  -\rho^{-1/2}(x) \cdot \varpi (x) \big).
	\end{split}{}
	\end{equation}
	This map is an embedding of the $\varphi$-phase space into the $\phi$-phase space with two important properties. It maps the mode functions $\{ g_k\}_k$ onto their pendants $\{f_k \}_k$, leaving the eigenvalue with respect to their respective kinetic operator invariant. And thus, due to its linearity, the image of any solution of the linearized equation of motion for $\varphi$ is again a solution of the linearized equations of motion for $\phi$. This implies in particular that the time evolutions of both theories are equivalent within the constraints imposed by the applicability of the linearized theories. \newline 
	These constraints on the applicability of the linearized theories translate to a restriction on the configurations which can be described in terms of the warp field. The inverse of \eqref{Redef} is given by 
	\begin{align}
	\varphi (t,x)= -\nu^{-1} \big(\sigma^\prime (\mu_m x) \big)^{-1} \cdot \phi (t,x).
	\end{align}
	However, in order for the linearized theory of $\varphi$ to apply, the condition \eqref{approx} has to be satisfied. This is only possible if $\phi$ is bounded by $|\phi| < b \nu \sigma^\prime (\mu_m x)  $, i.e. if it localized around the soliton's center. Inversely, on the linear level, the warp fields are only capable of describing such configurations.
	
	\subsection{Nonlinear embedding \& Applicability of the theory}
	The arguments of the last subsection can be generalized to the full, nonlinear theory. The identification of the warp field's zero mode with the collective coordinate naturally extends to the nonlinear level, so that it remains to lift the map between the non-zero modes to the nonlinear theory. This can be done by considering the map $R$, defined by
	\begin{equation}
	\begin{split}
	R: \mathcal{P} & \rightarrow \mathbb{P} \\ 
	\big( \varphi (x), \varpi (x) \big)  \mapsto \big( & R_\phi (\varphi)  (x) , R_\pi (\varphi,\varpi)  (x) \big), \label{EmbNL}
	\end{split}
	\end{equation}
	where
	\begin{equation}
	\begin{split}
	R_\phi (\varphi)  (x) \equiv & \nu \sigma (\mu_m x - \varphi (x) ) - \nu \sigma (\mu_m x ), \\  \label{p_vp}
	R_\pi (\varphi,\varpi) (x) \equiv &- \rho^{-1/2}(x,\varphi) \cdot \varpi (x).
	\end{split}
	\end{equation}
	This map, which assigns to every warp field configuration the fluctuation induced by it, is an embedding of the full phase space of the warp field, $\mathcal{P}$, into the one of general fluctuations around the soliton, $\mathbb{P}$. As it is clear that the equations of motion of a general fluctuation must also apply for the ones induced by the warp field, this embedding maps any solution of the $\varphi$-theory onto one of the $\phi$-theory. \newline 
	This map further reveals that the mode functions of $\varphi$ are no longer equivalent to the ones of $\phi$ on the non-linear level. Instead, a plane wave in the $\phi$-picture corresponds to a superposition of $\varphi$-waves and vice versa. \newline
	The inverse of this embedding can now be used to obtain a condition which configurations need to satisfy in order to be describable in terms of a warp field. The inverse of \eqref{EmbNL} is given by
	
	\begin{equation}
	\begin{split}
	R^{-1}:  \mathbb{P} \supset \mathcal{D}\big(R^{-1}\big) & \rightarrow \mathcal{P} \\ 
	\big( \phi (x), \pi (x) \big)  \mapsto \big( & R_\varphi^{-1} (\phi)  (x) , R_\varpi^{-1} (\phi,\pi)  (x) \big), \label{InvEmbNL}
	\end{split}
	\end{equation}
	with
	\begin{equation}
	\begin{split}
	R_\varphi^{-1} (\phi)  (x) \equiv &  -\sum_{n=1}^\infty \dfrac{\big(\sigma^{-1}  \big)^{(n)}(z)}{\nu^n n!}\bigg|_{z=\sigma (\mu_m x)} \phi^n (t,x), \\
	R_\varpi^{-1} (\phi,\pi)  (x) \equiv & - \rho^{1/2} \big(x,R_\varphi^{-1} (\phi) \big)\cdot \pi(x)  .
	\label{vpserInv}
	\end{split}
	\end{equation}
	Following the arguments above, given a fluctuation in $\mathcal{D}(R^{-1})$, this map allows to identify the warp field configuration causing said fluctuation. 
	The important question is now which kind of fluctuations form $D(R^{-1})$, i.e. which part of the phase space is covered by the theory of warp fields. This domain is constrained by the convergence of the series \eqref{vpserInv} or, equivalently, the invertibility of equation \eqref{p_vp}. From the latter one it is straightforward to derive a condition for $\phi$,
	\begin{align}
	\bigg|\frac{\phi (t,x)}{\nu} + \sigma (\mu_m x) \bigg| < 1. \label{Condi}
	\end{align}
	Recall that, for the sake of simplicity, the soliton is centered around $x=0$, i.e. the contributions of the zero modes are being ignored. The left hand side being equal to $1$ corresponds to the case $\varphi = \pm \infty$. \newline
	Hence, $\mathcal{D}(R^{-1})$ is given by  
	\begin{align}
	\mathcal{D}(R^{-1})= \{ (\phi (x), \pi (x)) \in \mathbb{P}| | \phi (x)/\nu + \sigma (\mu_m x) | < 1 \}.
	\end{align}{}
	As $\phi$ is in general time-dependent, it is possible that a certain configuration satisfies \eqref{Condi} within some time interval $T$, but ceases to do so for $t \nin T$. A physical interpretation of this property is given in the next subsection. \newline 
	Given a fixed time interval $T=[t_i, t_f]$, one can define $\mathcal{D}_T(R^{-1})$ as the set of configurations which remain within $\mathcal{D}(R^{-1})$ during $T$,
	\begin{equation}{}
	\begin{split}
	\mathcal{D}_T(R^{-1}) = \{ & (\phi (x), \pi (x)) \in \mathcal{D}(R^{-1})| (\phi (x), \pi (x)) = (\phi (t_i,x), \pi (t_i,x)) \\
	& \Rightarrow (\phi (t,x), \pi (t,x)) \in \mathcal{D}(R^{-1}) \forall t \in T\}.
	\end{split}
	\end{equation}{}
	Thus, within some fixed time interval $T$, all configurations which at $t=t_i$ are an element of $\mathcal{D}_T(R^{-1})$ can be described in terms of warp fields. \newline 
	The fact that only configurations localized around the soliton's center can be described in terms of warp fields is best understood from the construction of the warp field as an extension of the domain of the Goldstone field. The latter one is a result of the breaking of the translational symmetry. This symmetry is approximately restored in regions sufficiently far away from the center of the soliton, as the value of the field converges towards some constant value $\pm \nu$. It can therefore be expected that the physical significance of the warp field is biggest near the soliton's center, and vanishes sufficiently far away.
	
	\subsection{Incompleteness of the field space}
\begin{figure}
		\centering
		\includegraphics[trim = 112mm 83mm 131mm 85mm, clip, width=0.6\textwidth, height=0.5\textwidth]{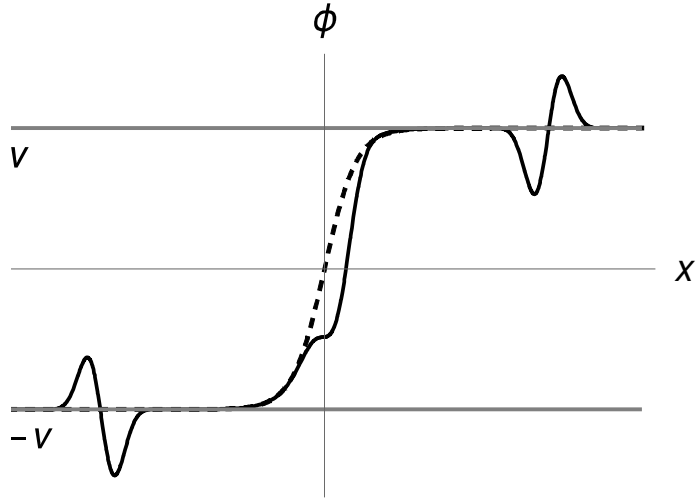}
		\caption{The soliton together with the wave package before, while and after passing through the soliton (solid line), compared with the unperturbed soliton (dotted line).}
		\label{Transition}
	\end{figure}

	The embedding of the theory of the warp field into the full theory of fluctuations has shown that it is possible for configurations to evolve into and out of the sector describable in terms of the warp field. In other words, the dynamical theory of the warp field is classically incomplete. While there exist situations in which such an incompleteness can be cured in the quantum theory (see \cite{Stefan}), this is not the case here. It is rather a necessary consequence of the fact that the warp field describes only a subset of the full theory. To understand the physical significance of this property, consider some localized wave package, propagating through the soliton in such a way that it is an element of $\mathcal{D}_T$ for some interval $T$ (Fig. \ref{Transition}). 
	Following the standard interpretation of the wave package as a particle, this process can be understood as said particle propagating through the soliton, being deformed while doing so and potentially picking up some phase shift. The notion of the warp field corresponds to an alternative perspective: While the fluctuation is localized in the same interval as the soliton, it can equivalently be understood as a warping of the latter. Thus, this process can be described as a wave package hitting the soliton and being continuously absorbed, causing the soliton to warp, a process described by the warp field. After the interval $T$ has passed, the soliton returns to an unexcited state by continuously emitting another wave package.  \newline
	This is in agreement with the picture developed in the previous sections: $\mathcal{D}_T$ is the set of configurations which can equivalently be understood as a warping of the soliton within the time interval $T$, and the warp field is nothing but an effective description for the dynamics of such configurations during $T$. \newline 
	The existence of such configurations is no surprise. In the context of a domain wall, the warp field can be understood as a perturbation in its thickness. However, it is well-known that such perturbations can decay into scalar particles \cite{Shifman}. The notion of the warp field now allows to give a clear characterization of this decay as the configuration reaching the boundary of the phase space of the warp field theory $\mathcal{P}$, i.e. $\varphi$ becoming divergent in some point $x_0$, which corresponds to $\sigma (\mu_m x_0 - \varphi (t,x_0))=\pm 1$, i.e. $\Phi(x_0)= \pm\nu$.

	\subsection{Warp fields in different regimes}
	The region in which the contribution of the warp field to the full dynamical theory is of relevance depends strongly on the localization of the soliton, and therefore indirectly also on the parameters of the underlying theory. In the simple setting discussed in this article, these parameters are the asymptotic values the soliton converges towards, the coupling constant and the mass of the scalar field. In general only two out of these three parameters are independent. Taking as an example the case of the $\Phi^4$-theory, these coefficients are connected by the relation $m_\Phi \propto \nu \sqrt{\lambda}$. \newline
	As the chosen parametrization does not explicitly depend on the coupling constant $\lambda$, it is most convenient to choose the vacua $\pm \nu$ and the mass of the field $m_\Phi$ as independent parameters. It is also important to recall that the warp field $\varphi$ is obtained from the collective coordinate, i.e. the Goldstone field, not only by extending its domain, but by also by multiplying it with a factor of $\mu_m \propto m_\Phi$, so that $\varphi \propto m_\Phi$.\newline
	Consider first the case $m_\Phi \to 0$, implying that also $\mu_m \to 0$. In this limit, the soliton is widely outstretched, while $\varphi \to 0$. This is in agreement with the origin of the warp field in the breaking of translational symmetry in the presence of the soliton. As the soliton spreads out in this limit, the same is true for the region in which the symmetry is locally broken. However, at the same time the soliton flattens out, so that the symmetry breaking becomes less distinct. Note that, given a fixed value of $\nu$, this limit is equivalent to the one of a weak interaction. \newline
	In the limit $m_\Phi \to \infty$, when also $\mu_m \to \infty$, the opposite can be observed. With the soliton becoming increasingly localized, the region affected by the warp field shrinks until the latter one is fully confined to the soliton, which can with increasing precision be approximated as a point particle. Therefore, in the limit of a heavy field or, equivalently, a strong interaction with fixed $\nu$, the warp field becomes identical with the collective coordinate. This can also be observed from \eqref{Condi}. In the limit of a soliton which is fully localized to a point, \eqref{Condi} can only be solved by $\phi=0$, i.e. the non-zero mode contributions to the warp field have to disappear. This also resolves the issue that taking this limit seems to imply $\varphi \to \infty$. As the non-zero mode contributions disappear in this limit, only the terms representing the collective coordinate remain. Their prefactor $\mu_m$ factorizes out inside the argument of the soliton, and outside integrates with the weight factor to the soliton's mass. \newline
	The explicit $\nu$-dependence is easily understood: As the energy density of the soliton scales as $\nu^2$, the same is true for all physical effects caused by the warp field.

	\subsection{Geometric perspective}
	In the presence of gravity, the translational Goldstone fields are eaten up by the metric, as they can be understood as the action of some diffeomorphism on the soliton \cite{DvaliShifman}. As this letter statement is also true for the warp field, it seems only natural that it can also be absorbed in a similar way. \newline
	To see that this is indeed the case, consider the coordinates \[\big(\xi^0 (t,x),\xi^1 (t,x)\big)  =\big(t, x - \mu_m^{-1}\varphi (t,x) \big).\] In terms of these coordinates, the action of the soliton can be rewritten as
	
	\begin{equation}
	\begin{split}
	S[\Phi] =  \intd ^2 \xi \sqrt{- g_\varphi } \ \bigg(& \dfrac{1}{2}  g^{\mu \nu}_\varphi (\xi) \partial_\mu \Phi_s(\xi) \partial_\nu \Phi_s (\xi) - V(\Phi_s (\xi (t,x) )\bigg),
	\end{split}
	\end{equation}
	where the effective metric is given by
	\begin{equation}
	g_{\mu \nu}^\varphi (\xi)= \eta_{\alpha \beta} \frac{\partial x ^\alpha}{\partial \xi^\mu} (\xi) \frac{\partial x ^\beta}{\partial \xi^\nu} (\xi) , \label{g_vp}
	\end{equation}
	whose determinant is as usual denoted by $g_\varphi$.  \newline
	In this new action, all the $\varphi$s are combined into $g_{\mu \nu}^\varphi$. Hence, it is reasonable to ask whether this can be taken one step further, describing the dynamics completely by $g_{\mu \nu}^\varphi$. Formally this corresponds to promoting $g_{\mu \nu}^\varphi$ to a field, which will be denoted by $g_{\mu \nu}$ to point out that it is no longer considered a function of $\varphi$. When doing so, the information about its structure \eqref{g_vp} has to be implemented via a Lagrange multiplicator $\Lambda$. In 1+1 dimensions, \eqref{g_vp} is equivalent to $R=0$, where $R$ is the usual Ricci scalar \cite{Weinberg}. Hence, the dynamics of the warp field is fully encoded in the action
	
	\begin{equation}
	\begin{split}
	S[\Phi_s , g ]= \intd ^2 \xi \ \sqrt{-g} ( \mathcal{L}_S + \Lambda R ),
	\end{split}
	\end{equation}
	where $\mathcal{L}_s$ denotes the action of the soliton on the spacetime with metric $g$. The term multiplied by the Lagrange multiplier is nothing but the Einstein-Hilbert action, formally providing a kinetic term for the metric. This action gives rise to the equation of motion for $g_{\mu \nu}$,
	
	\begin{equation}
	\begin{split}
	T^{\mu \nu}=&  2( \Lambda G^{\mu \nu} -   \nabla ^{\nu} \partial^\mu \Lambda  + g^{\mu \nu}  \Box_g \Lambda ),  \label{Einstein}
	\end{split}
	\end{equation}
	where $T^{\mu \nu}$ denotes the energy momentum tensor of the scalar field, $T^{\mu \nu} = \frac{\delta \sqrt{-g} \mathcal{L}_s}{\delta g_{\mu \nu}}$. Equation \eqref{Einstein} together with the constraint $R=0$ fully determines the dynamics of the theory, parametrized by $g_{\mu \nu}$ and $\Lambda$. The constraint implies in particular that the kinetic term for $g_{\mu \nu}$, $G^{\mu \nu}$, is identical to zero, so that the metric is no propagating degree of freedom. Instead, this role is taken by the Lagrange multiplier $\Lambda$. \newline
	More precisely, $\Lambda$ acts as a dynamical degree of freedom sourced by the soliton's energy density. This can be seen by inserting the energy-momentum tensor of the scalar field into \eqref{Einstein} and contracting both sides with $g_{\mu \nu}$, yielding
	\begin{equation}
	\begin{split}
	\Box_g \Lambda = \text{tr}_g(T)= V(\Phi_s (\xi)) = \frac{1}{2} \epsilon_s(\xi). \label{NiceEq}
	\end{split}
	\end{equation}
	Just as in the $\varphi$-theory, the soliton enters only via its energy density. \newline 
	This equation corresponds to the action
	\begin{equation}
	\begin{split}
	\Tilde{S}=\intd^2\xi \sqrt{-g} \ \bigg(\frac{1}{2} \partial_\mu \Lambda (\xi) \partial^\mu \Lambda (\xi)+ \Lambda (\xi) \epsilon_s(\xi) \bigg).
	\end{split}
	\end{equation}
	This action - which is of an even simpler form as \eqref{Actionf} - describes the full dynamics of the warp field sector for which \eqref{g_vp} is non-degenerate.

	\section{Warp fields in higher dimensions}
	
	\subsection{General aspects}
	In the $(1+1)$-dimensional case, the warp field gives rise to the full dynamics of the fluctuation $\phi$ near the soliton's center, a feature based on the coincidence that the number of fields matches the number of broken translational symmetries. This is in general not the case, as for a generic theory, the number of fields necessary to form a soliton-like configuration is in no way related to the number of spatial dimensions\footnote{Except for the restrictions imposed by Derrick's theorem of course, which are not relevant here.}. It is nevertheless in any case possible to equip a given theory with warp fields by simply promoting the collective coordinates corresponding to the soliton's position to fields. The crucial question is then whether or not the theory of these warp fields covers a subsector of the dynamics of all fields of the underlying theory. If this is the case, their theory will be called \textit{sufficient}, otherwise \textit{insufficient}. \newline 
	To formalize this distinction, let $\{ \varphi^a\}_a$ denote a set of warp fields. Let further denote $\{ \phi_i \}_i $ a set of real fields parametrizing the fluctuations around the soliton giving rise to these warp fields. The theory of these warp fields is sufficient, if there exists some invertible function $\mathcal{R}$ respecting the equations of motion of both $\{ \varphi^a\}_a$ and $\{ \phi_i \}_i $ s.t.
	\begin{equation}
	\begin{split}
	\phi_i (t,x) = \mathcal{R}_i[\varphi](t,x).
	\end{split}{}
	\end{equation}{}
	Considering only the linearized theories, this corresponds to the existence of some invertible matrix $R_{is}$ s.t.
	\begin{equation}
	\begin{split}
	\phi_i (t,x) =R_{ia}(x) \varphi^a (t,x).
	\end{split}{}
	\end{equation}{}
	The case of a sufficient theory of warp fields is straightforward and discussed in great detail in section 3.2. It is then illustrated using the important example of the Skyrme model in section 3.3. \newline 
	In the case of an insufficient theory, some additional effort is necessary. Any attempt to use such a theory to describe the dynamics of the full theory is spoiled by the insufficient number of independent fields. Recall that the construction of the theory of warp fields sets out from the soliton's moduli space. The crucial step towards an extension of an insufficient theory of warp fields lies now in the observation that for most higher-dimensional theories, there exist zero modes besides the ones corresponding to spatial translations. These additional zero modes represent e.g. (gauge-)rotations, and as they are massless they are readily available in the same regime as the collective coordinates. Thus, they serve as natural candidates to provide the missing degrees of freedom. A theory of warp fields, for which it is possible to compensate for the missing degrees of freedom by taking into account additional zero modes will be called \textit{extendable}. \newline 
	This is discussed in section 3.4, and demonstrated on the example of the abelian Higgs model in $2+1$ dimensions in section 3.5.

	\subsection{Sufficient theories}
	In the case of sufficient theories, the warp fields suffice to capture a subsector of the dynamics of all the fields involved in the formation of the soliton, and all configurations localized around the soliton's center can be fully described in terms of their theory. The first step towards the construction of their Lagrangian lies in the observation that the warp fields enter their underlying theory only in the combination
	\begin{align}
	\xi^i \equiv \mu_m x^i - \varphi^i(t,x)
	\end{align}{}
	inside the argument of the soliton. Let now $\Phi$ denote an arbitrary field contributing to the soliton. Introducing the warp fields transforms this field and its partial derivative as
	\begin{equation}
	\begin{split}
	\Phi_s (\mu_m x) &\to \Phi_s (\xi ), \\ 
	\partial_\mu \Phi_s (\mu_m x) &\to \partial_a \Phi_s (\xi) \big( \mu_m \delta^a_\mu - \varphi^a_{,\mu}\big),
	\end{split}{}
	\end{equation}{}
	where $\partial_a$ denotes the derivative with respect to the dimensionless variable $\xi^a$. \newline 
	Thus, for a theory containing up to $N$ derivatives of its field content, the most general Lagrangian in terms of the warp field is of the form
	\begin{equation}
	\begin{split}
	\mathcal{L}=& g( \xi) + \sum_{n=1}^N C_{a_1 \dots a_n}^{\mu_1 \dots \mu_n} (\xi) \xi^{a_1}_{,\mu_1} \dots \xi^{a_n}_{,\mu_n}. \label{LagrAllg}
	\end{split}{}
	\end{equation}{}
	All the information about the underlying theory enters only via the coefficients $C_{a_1 \dots a_n}^{\mu_1 \dots \mu_n}(\xi)$ as well as the function $g$ . Their contraction with the factors of $\xi$ implies that the coefficients $C_{a_1 \dots a_n}^{\mu_1 \dots \mu_n}(\xi)$ are invariant under pairwise exchange of two upper and their corresponding lower indices, e.g.
	\begin{equation}
	C_{a_1 a_2 \dots a_n}^{\mu_1 \mu_2 \dots \mu_n} = C_{a_2 a_1 \dots a_n}^{\mu_2 \mu_1 \dots \mu_n}.
	\end{equation}{}
	At this point, this action still contains a large number of unknowns, in particular the number of terms in \eqref{LagrAllg}. This is where the warp fields' relation to the collective coordinates becomes important, as it almost completely determines the theory's structure. \newline 
	In the limit $\varphi \to 0$, \eqref{LagrAllg} has to reduce to the Lagrangian of the unperturbed soliton. In the limit $\nabla \varphi \to 0$, the remaining terms have to reduce to the effective theory in terms of the collective coordinates. Thus, $N$ is given be the highest order of time derivatives found in the low-energy theory. \newline 
	To see how this constrains the theory, consider the most important case of an action which is quadratic in the collective coordinates' velocities, 
	\begin{equation}
	L[z]= \frac{1}{2} M_{ab} \dot{z}^a \dot{z}^b, \label{LagrCC}
	\end{equation}{}
	where $M_{ab}$ is the usual moduli space metric. From here, following the arguments above leads to a Lagrangian of the form
	\begin{equation}
	\begin{split}
	\mathcal{L}[\varphi]=& g(\xi)+  C_{a}^{\mu} (\xi) \big(\mu_m \delta^a_\mu - \varphi^a_{,\mu}\big)+ C_{ab}^{\mu \nu} (\xi) \big(\mu_m \delta^a_\mu - \varphi^a_{,\mu}\big)\big(\mu_m \delta^b_\nu - \varphi^b_{,\nu}\big) = \\ 
	=& \big( g(\xi)+ \mu_m  C_{a}^{a} (\xi) + \mu_m^2 C_{ab}^{a b} (\xi) \big)-\big( 2 \mu_m C^{\mu b}_{a b} (\xi) + C_{a}^{\mu} (\xi)\big) \varphi^a_{, \mu}  + C_{ab}^{\mu \nu} (\xi) \varphi^a_{,\mu} \varphi^b_{,\nu}.
	\end{split}{}
	\end{equation}{}
	The limit $\varphi \to 0$ implies that the terms in the first bracket have to coincide with the soliton's energy density evaluated at $\xi$. Their scaling with $\mu_m$
	allows to identify them with the soliton's energy due to (self-)interactions ($g$), some kinetic coupling ($C_{a}^{a}$), e.g. involving some gauge field, and the gradient energy ($C_{ab}^{a b}$),
	\begin{equation}
	\begin{split}
	E_{\text{grad}}=&- \intd^3 x \ \mu_m^2 C_{ab}^{ab} (\mu_m x) ,\\ 
	E_{\text{kc}}=&- \intd^3 x \ \mu_m C_{a}^{a} (\mu_m x) , \\
	E_{\text{int}}=&- \intd^3 x \ g (\mu_m x) . \label{EFormeln}
	\end{split}{}
	\end{equation}{}
	The limit $ \nabla \varphi \to 0$ and the absence of a term linear in $\dot{z}$ in \eqref{LagrCC} imply that 
	\begin{equation}
	2 \mu_m C^{0 b}_{a b} (\mu_m x) + C_{a}^{0} (\mu_m  x) =0, \label{NoLin}
	\end{equation}{}
	i.e. the term linear in $\dot{\varphi}$ disappears. Finally, the coefficients of the term quadratic in derivatives of $\varphi$ have to reduce to the moduli space metric when integrated over space,
	\begin{equation}
	\intd^3 x \ C_{ab}^{00} (\mu_m x) = \frac{1}{2} M_{ab}. 
	\end{equation}{}
	This can be simplified even further if one can safely assume that the derivatives of the underlying theory's kinetic term are contracted using the metric $\eta^{\mu \nu}$. If this is the case, $C_{ab}^{\mu \nu}$ can be rewritten as $\eta^{\mu \nu} \rho_{ab}$, where $\rho$ acts as a density of the moduli space metric,
	\begin{equation}
	\intd^3 x \ \rho_{ab} (\mu_m x) = \frac{1}{2} M_{ab}. 
	\end{equation}{}
	The fact that $C_{ab}^{a b}$ corresponds to the soliton's gradient energy can in this case be translated to a relation for $\rho$, 
	\begin{equation}
	\intd^3 x \ \eta^{ab} \rho_{ab} (\mu_m x) = - E_{\text{grad}}. 
	\end{equation}{}
	This additional assumption also implies immediately that $C^{0 b}_{a b} =0$, so that $C_{a}^{0} =0$ due to \eqref{NoLin}. \newline 
	Summarizing all the previous arguments, the final Lagrangian is of the form
	\begin{equation}
	\begin{split}
	\mathcal{L}[\varphi]=&- \epsilon_s (\xi) + \rho_{ab} (\xi) \varphi^a_{,\mu} \varphi^{b,\mu} + \kappa_a^b (\xi) \varphi^a_{,b}  . \label{LWarpdfin}
	\end{split}{}
	\end{equation}{}
	Given that the theory of a certain soliton's moduli is quadratic in their derivatives and assuming that the underlying theory's kinetic term is constructed using the metric, this simple structure is universal. It holds in full non-linearity, and the only quantities sensitive to the details of the underlying theory are the functions $\rho_{ab}, \kappa_a^b$ and $\epsilon_s$. Decomposing these functions in terms of the $C$s allows to relate them to physical properties of the underlying soliton.\newline
	One of the warp fields' central features found in $1+1$ dimensions is the localization around the soliton's center. It is easy to see that the same is true in higher dimensions. The convergence of the integrals \eqref{EFormeln} implies that $\rho_{ab}$ (or $C_{ab}^{\mu \nu}$, respectively), which takes the role of the weight function, has to decay at least as $|x|^{-2}$ for $|x| \to \infty$. Thus, all the effects arising from a localized weight function can be expected to occur also in this higher-dimensional setting. \newline 
	
	\subsection{Warp fields of the Skyrme model as a sufficient theory}
	The Skyrme model is a particularly instructive example for a sufficient theory due to its higher derivative terms, which require a modification of the discussion given in the last subsection. As an effective field theory, this model contains terms of fourth order of the fields' derivatives, implying that also terms of up to fourth order in $\xi^a_{,\mu}$ have to be taken into account. To deduce the dynamics around the Skyrmion in terms of warp fields up to some background functions, only two pieces of information are necessary. First, that there are three independent fields involved, matching the number of spatial dimensions and thus broken translational symmetries. Second, that the action of the collective coordinates contains terms of quadratic and quartic order in their time derivatives. Assuming again that $\eta^{\mu \nu}$ is the only tensor used to contract the indices of the underlying theory's partial derivatives, the most general Lagrangian compatible with the properties of the warp fields is of the form
	
	\begin{equation}
	\begin{split}
	\mathcal{L}=& \frac{1}{2} \rho_{ab}(\xi) \partial_\mu \varphi^a \partial^\mu \varphi^b  +(\Lambda^4)_{abcd}(\xi) \partial_\mu \varphi^a \partial^\mu \varphi^b \partial_\nu \varphi^c \partial^\nu \varphi^d - \epsilon_s (\xi) + \\ 
	&+  (\Lambda^1)^i_a(\xi) \partial_i \varphi^a + (\Lambda^2)^{ia}_{jb}(\xi) \partial_i \varphi^j \partial_a \varphi^b + (\Lambda^3)_{abc}^i(\xi)  \partial_\mu \varphi^a \partial^\mu \varphi^b \partial_i \varphi^c. \label{LagrWarpSky}
	\end{split}{}
	\end{equation}{}
	As the Skyrme-model is well-known\footnote{See e.g. \cite{Skyrme},\cite{Skyrme0} for a review.}, the coefficients can in this case be determined by a direct computation. \newline 
	In $3+1$ dimensions, the dynamics of the Skyrme model can be described using three $SU(2)$-phases representing pion fields,
	\begin{align*}
	U(t,x)=\exp{\bigg(\frac{i}{f_\pi} \pi^a (t,x) \tau_a \bigg)},
	\end{align*}{}
	where $f_\pi$ denotes the pion decay constant and the $\{\tau_a\}_a$ are a set of orthonormal $SU(2)$-generators, usually the Pauli matrices. Their leading order Lagrangian including the soliton-generating interaction term is given by
	\begin{equation}
	\begin{split}
	\mathcal{L} &= \dfrac{f_\pi^2}{4} \mathrm{tr} \big( \partial_\mu U \partial^\mu U^\dagger \big) +  \dfrac{1}{32 e^2}  \mathrm{tr}\big( [\partial_\mu U , \partial_\nu U^\dagger] [\partial^\mu U , \partial^\nu U^\dagger] \big)  \label{SkyrmAct}
	\end{split}
	\end{equation}
	The parameter $e$ is the \textit{Skyrme constant}. This theory gives rise to topological solitons, so-called Skyrmions, which are of the form
	\begin{align*}
	U_s(r)=\exp \big( i F(r) \hat{x}^a \tau_a \big),
	\end{align*}{}
	where $\hat{x}^a=\frac{x^a}{r}$ and $F(r)$ is some function minimizing the configuration's energy. On the linearized level, the fluctuations around this soliton are elements of the Lie-algebra su$(2)$, and can thus be expressed in terms of the generators $\{\tau^a \}_a$ as
	\begin{align}
	\delta U (t,x)=\delta U^a(t,x) i \tau_a. \label{FlucSky}
	\end{align}{}
	\newline 
	Note that for convenience, the factor of $\frac{1}{f_\pi}$ has been absorbed into $\delta U$ here. The transformation from fluctuations to warp fields and back can now be determined explicitly. Introducing a warp field in the usual way transforms the Skyrmion to first order as
	\begin{align}
	U_s(r)\to U_s(r) + i \tau_a \varphi^j (t,x) \mu_m^{-1} \bigg[ \hat{x}^a \hat{x}^j \bigg(\frac{F(r)}{r}-F^\prime (r) \bigg) - \delta^a_j \frac{F(r)}{r} \bigg) \bigg].
	\end{align}{}
	Comparing this with \eqref{FlucSky} immediately leads to the transformation 
\begin{align}
	\delta U^a(t,x) = \mu_m^{-1} M^a_j (x) \varphi^j (t,x), 
\end{align}{}
	where
\begin{align}
	M^a_j (x) = \hat{x}^a \hat{x}^j \bigg(\frac{F(r)}{r}-F^\prime (r) \bigg) - \frac{F(r)}{r} \delta^a_j. 
\end{align}{}
	This map is invertible, and its inverse is given by
	\begin{align}
	(M^{-1})^j_a (x) = \hat{x}^a \hat{x}^j \bigg(\frac{r}{F(r)}-\frac{1}{F^\prime (r)} \bigg) - \frac{r}{F(r)} \delta^j_a. 
	\end{align}{}
	As $F(r) \to 0$ for $r \to \infty$, $M^{-1}$ diverges for large $r$. Similarly as in the $(1+1)$-dimensional case, this together with the conditions necessary for the validity of the linearized theory restricts the applicability of the warp fields to configurations which decay sufficiently fast, i.e. are localized around the Skyrmion's center, in agreement with the general argument given in the last subsection. \newline 
	Now that it is clear that the theory of fluctuations around the Skyrmion can in fact be formulated in terms of warp fields, it remains to determine the coefficients in \eqref{LagrWarpSky}. Inserting the warped Skyrmion into \eqref{SkyrmAct}, one finds
	\begin{equation}
	\begin{split}
	\rho_{ab} (\xi) =& \dfrac{f_\pi^2}{4} \mathrm{tr} \big( \partial_a U_s (\xi) \partial_b U_s^\dagger (\xi) \big) +  \dfrac{1}{32 e^2}  \mathrm{tr}\big( [\partial_a U_s (\xi), \partial_j U_s^\dagger (\xi)] [\partial_b U_s (\xi) \partial^j U_s^\dagger (\xi)] + \\ 
	&+[\partial_j U_s (\xi), \partial_a U_s^\dagger (\xi)] [\partial^j U_s (\xi) \partial_b U_s^\dagger (\xi)] \big) \\
	(\Lambda^4)_{abcd}(\xi) =&\frac{1}{32 e^2} \mathrm{tr}\big([\partial_a U_s (\xi), \partial_c U_s^\dagger (\xi)] [\partial_b U_s (\xi) \partial_d U_s^\dagger (\xi)] \big) \\ 
	(\Lambda^1)^b_a(\xi)=&  \mu_m \frac{f_\pi^2}{4} \mathrm{tr}\big( \partial_{(a} U_s (\xi) \partial_{b)} U_s^\dagger (\xi) \big)+ \\ 
	&+\frac{\mu_m^3}{32e^2}\mathrm{tr}\big( [\partial_{(a} U_s (\xi), \partial^j U_s^\dagger (\xi)] [\partial_{b)} U_s (\xi) \partial_j U_s^\dagger (\xi)] + \\ 
	&+[\partial_j U_s (\xi), \partial_{(a} U_s^\dagger (\xi)] [\partial^j U_s (\xi) \partial_{b)} U_s^\dagger (\xi)] \big) \\ 
	(\Lambda^2)^{ia}_{jb}(\xi)=& \frac{\mu_m^2}{16 e^2} \mathrm{tr}\big([\partial_j U_s (\xi), \partial_b U_s^\dagger (\xi)] [\partial_i U_s (\xi) \partial_a U_s^\dagger (\xi)] + \\ 
	&+[\partial_j U_s (\xi), \partial_a U_s^\dagger (\xi)] [\partial_i U_s (\xi) \partial_b U_s^\dagger (\xi)] \big) \\ 
	(\Lambda^3)_{abc}^i(\xi)=& \frac{\mu_m}{16 e^2}  \mathrm{tr}\big([\partial_a U_s (\xi), \partial_c U_s^\dagger (\xi)] [\partial_b U_s (\xi) \partial^j U_s^\dagger (\xi)] + \\ 
	&+ [\partial_c U_s (\xi), \partial_a U_s^\dagger (\xi)] [\partial^j U_s (\xi) \partial_b U_s^\dagger (\xi)] \big)
	\end{split}
	\end{equation}

\subsection{Extendable theories}
For most theories giving rise to solitons, the number of independent fields is larger than the number of broken translational symmetries, so that the collective coordinates corresponding to these symmetries fail to give rise to a sufficient theory of warp fields. It is now important to note that the central benefits of constructing the theory of warp fields from these collective coordinates, namely their universality as well as their relevance for the low-energy dynamics, are not unique to this particular setup. They are also shared by any massless excitation, i.e. any zero mode. As the construction of the warp fields already relies on information gained from this sector, utilizing potentially existing, additional zero modes to compensate for the lacking degrees of freedom appears as a natural path to extend the applicability of the construction of warp fields. \newline
In the following, these additional zero modes will be denoted by $\{\theta^a \}_a$, and the fields obtained from them by $\{\vartheta^a (t,x)\}_a$. As these zero modes usually correspond to (gauge)-rotations and thus promoting them to fields can be visualized as a twirling of the configuration, these fields will be referred to as \textit{twirl fields}. Introducing them into the action in the same way as the warp fields modifies the partial derivatives of the fields involved in the formation of the soliton as
\begin{equation}
	\partial_\mu \Phi_s (x) \to \partial_a\Phi_s (\xi, \vartheta) (\mu_m \delta_\mu^a - \varphi^a_{,\mu}) + \partial_{\vartheta^a}\Phi_s (\xi, \vartheta) \vartheta^a_{,\mu}.
\end{equation}
Recall that the theory of pure warp fields was constructed from a suitable number of $\xi^a_{, \mu}$'s, with the highest number being linked to the highest order of derivatives of the collective coordinates as well as some coefficients, which were functions of $\xi$ and contained the information about the underlying theory. In this picture, the dynamics of the twirl fields can be taken into account by supplementing the Lagrangian with terms containing suitable orders of $\vartheta^a_{,\mu}$ as well as terms mixing them with the $\xi^a_{, \mu}$'s.\newline
Consider now again the case of a theory whose low-energy theory is only of second order in time-derivatives. Independent of the field content or the exact form of the soliton, the most general Lagrangian in terms of both warp and twirl fields is of the form
\begin{equation}
\begin{split}
	\mathcal{L}[\varphi]=& g(\xi)+  C_{a}^{\mu} (\xi) \big(\mu_m \delta^a_\mu - \varphi^a_{,\mu}\big)+ C_{ab}^{\mu \nu} (\xi) \big(\mu_m \delta^a_\mu - \varphi^a_{,\mu}\big)\big(\mu_m \delta^b_\nu - \varphi^b_{,\nu}\big) + \\
	&+D_{ab}^{\mu \nu} (\xi,\vartheta) \partial_\mu \vartheta^a \partial_\nu \vartheta^b + D_a^\mu (\xi,\vartheta) \partial_\mu \vartheta^a + M_{ab}^{\mu \nu} (\xi,\vartheta) \big(\mu_m \delta^a_\mu - \varphi^a_{,\mu}\big) \partial_\nu \vartheta^b. \label{Righthere}
\end{split}{}
\end{equation}{}
As the twirl fields parametrize a symmetry, they have to disappear from the theory in the limit $\vartheta^a_{,\mu} \to 0$. Thus, the coefficients in the first line of \eqref{Righthere} have to be independent of $\vartheta$. The coefficients $D_{ab}^{\mu \nu}$ can be linked to the low-energy theory's field space metric in the same way as $C_{ab}^{\mu \nu}$, and the restriction to second-order terms in the low-energy theory implies that the terms linear in $\dot{\vartheta}$ vanish. Assuming further that the partial derivatives making up the kinetic term are contracted via the metric, the general Lagrangian \eqref{Righthere} can be brought to its final form,
\begin{equation}
	\begin{split}
	\mathcal{L}[\varphi,\vartheta]=&- \epsilon_s (\xi)+ \kappa_a^b (\xi) \varphi^a_{,b}  + \rho_{ab} (\xi) \varphi^a_{,\mu} \varphi^{b,\mu}+ \\ 
	&+\omega_{ab} (\xi, \vartheta)  \vartheta^a_{,\mu}  \vartheta^{b,\mu} + K_a^j (\xi, \vartheta) \vartheta^a_{,j} + L_{ab}^{\mu \nu} (\xi, \vartheta) \varphi^a_{,\mu}  \vartheta^b_{,\nu}. \label{LagrWarpTwist}
	\end{split}{}
\end{equation}{}
Just as in the case of a theory constructed from warp fields only, every theory can be supplemented with twirl fields. The crucial question is now whether or not this procedure generates enough degrees of freedom to capture the dynamics of all fields involved in the formation of the soliton. A priori, there is no guarantee that this is the case for any theory. Nevertheless, this approach strongly extends the applicability of this article's philosophy. One example therefore is presented in the next subsection.

\subsection{Warp fields of the abelian Higgs model in 2+1 dimensions as an extendable theory}
The abelian Higgs model in $2+1$ dimensions, which gives rise to the Abrikosov-Nielsen-Olesen vortex \cite{Shifman}, \cite{Nielsen}, serves as an important example for an extendable theory of warp fields, as its discussion can also serve as a showcase on how the challenges arising from gauge redundancies in the context of warp fields can be tackled. \newline 
Using the strategy given in the last subsection, the construction of a theory of warp and twirl fields around the vortex is straightforward once suitable zero modes have been identified. It is recapitulated within this section that the vortex provides three zero modes, while the underlying theory contains three physical degrees of freedom. The main challenge in the context of gauge theories lies however in the matching of these degrees of freedom. Due to the theory's gauge invariance, the number of fields forming the theory of interest is larger than the number of physical degrees of freedom. This prevents a straightforward application of the strategy used to relate the theory of warp fields to the full theory used in the previous settings, as the relation between warp fields and fluctuations is no longer invertible. A simple solution to this problem would be to eliminate the redundant fields using a sufficient number of gauge transformations. Unfortunately, on the level of the Lagrangian, this is not always possible. Take as an example the temporal gauge, which eliminates $A_0$. Any gauge transformation aimed at eliminating another field would have to be generated by a gauge function involving these fields, which are in general functions of time, so that it would be inconsistent with the temporal gauge. This changes however on the level of the equations of motion, as these can in certain cases be used to eliminate the time-dependence of the gauge function. This is precisely the case in the example discussed within this section, and will be used as follows: After a quick overview over the well-known theory of fluctuations around the vortex, a particularly useful gauge is presented, which eliminates all redundancies on the level of the equations of motion. This reduces the system to three independent fields, each of them subject to an independent equation of motion. \newline
Afterwards, the strategy presented in the last subsection is used to construct a theory of warp and twirl fields around the vortex. This will also result in three independent fields, whose dynamics is determined by three equations of motion. The fluctuations induced by these fields span a three-dimensional subspace of the four-dimensional field space of the (off-shell) fluctuations. This connection between the two theories can then be used to show that the equations of motions of the fluctuations together with an additional, on-shell gauge condition imply the ones of the warp fields and vice versa, thereby showing the equivalence of the two approaches up to the usual limitations arising from the use of warp fields. For simplicity, all of this is done for the linearized theories only. \newline 
The first step of this discussion is to recall the central properties of the ANO vortex. Due to the static nature of the soliton underlying the following discussion, it is convenient to work in the temporal gauge, $A_0=0$, which can be imposed on the level of the Lagrangian. This reduces the number of real, independent fields to four, two of which are given by the spatial components of the gauge field $A_i$. The other two are contained within the complex scalar field, and can be made manifest via the parametrization
\begin{equation}
    \Phi (t,x)= \frac{1}{\sqrt{2}} \rho(t,x) e^{i q \lambda (t,x)},
\end{equation}{}
where $q$ denotes the $U(1)$-charge appearing in the covariant derivative, $D_\mu = \partial_\mu + i q A_\mu$. \newline
In terms of these fields and the gauge-invariant quantity $J_i=A_i+\partial_i \lambda$, the Lagrangian is given by
\begin{equation}
    \mathcal{L}= \frac{1}{2} \partial_\mu \rho \partial^\mu \rho +\frac{q^2 \rho^2}{2} \dot{\lambda}^2+\frac{1}{2} \dot{A_i}\dot{A_i}- \frac{1}{4} F_{ij}F_{ij}-\frac{q^2 \rho^2}{2} J_i J_i - V(\rho), \label{LagrHiggs}
\end{equation}{}
where $V$ denotes a potential whose vacua are given by $\rho = \pm \nu$. In these conventions, the vortex solution of winding number 1 is of the form
\begin{equation}
    \begin{split}
        \rho_s=\nu \sigma (r), \ 
        \lambda_s= \frac{\alpha}{q}, \ 
         A_i^s = - \partial_i \alpha_s \big(1-f(r)\big)= \frac{1}{q} \epsilon_{ij} \frac{x_j}{r} \big(1-f(r)\big), \label{Vortex}
    \end{split}{}
\end{equation}{}
where $\alpha$ denotes the spatial angle \cite{Shifman}, \cite{Nielsen}. Denoting the fluctuations around $\rho_s$, $\lambda_s$ and $A_i^s$ by $r$, $l$ and $a_i$ respectively, the leading order Lagrangian of these fields can be obtained by a straightforward computation: 
\begin{equation}
    \mathcal{L}^{(2)}= \frac{1}{2}\partial_\mu r \partial^\mu r- \frac{q^2}{2} J_i^s J_i^s r^2 - \frac{1}{2} V^{\prime \prime }(\rho_s) r^2 + \frac{q^2 \rho_s^2}{2} \dot{l}^2 + \frac{1}{2}\dot{a}_i\dot{a}_i - \frac{1}{4} f_{ij} f_{ij} - 2 q^2 J_i^s \rho_s j_i r - \frac{q^2 \rho_s^2}{2} j_i j_i,
\end{equation}{}
where $f_{ij}= \partial_i a_j - \partial_j a_i$. This Lagrangian corresponds to the following equations of motion:
\begin{equation}
    \begin{split}
        \Box r=& - \big( q^2 J_i^s J_i^s +V^{\prime \prime}(\rho_s) \big)r - 2 q^2 J_i^s \rho_s j_i, \\ 
        \ddot{l}=& \rho_s^{-2} \partial_i \big( \rho_s^2 j_i +2 \rho_s J_i^s r \big) , \\ 
        \ddot{a_i} =& \partial_m f_{mi} - 2 q^2 J_i^s \rho_s r - q^2 \rho_s^2 j_i. \label{eomHiggs}
    \end{split}{}
\end{equation}{}
The system further underlies the constraint 
\begin{equation}
    \begin{split}
    \partial_i \dot{A_i} =- q ^2 \rho^2 \dot{\lambda}, \label{ConsHam}
\end{split}{}
\end{equation}{}
which can be obtained either from the equation of motion for $A_0$, or alternatively from the constraint that its canonical momentum vanishes for all times \cite{constrHam}. Consider now the gauge function
\begin{equation}
    \gamma = - \frac{1}{\Delta}\big(\partial_i A_i - G[\rho, \alpha]\big),
\end{equation}
where $G$ is the solution of the differential equation $\dot{G}=- q^2 \rho^2 \dot{\lambda}$. The constraint (or the equation of motion for $A_0$ respectively) \eqref{ConsHam} now implies that $\dot{\gamma}=0$, so that the gauge transformation generated by this function is consistent with the already imposed temporal gauge. Performing the gauge transformation generated by this function, $A_i \to A_i + \partial_i \gamma$ and $\lambda \to \lambda - \gamma$ leads to the identity
\begin{equation}
    \partial_i A_i = G. \label{gaugeCond}
\end{equation}{}
When considering fluctuations around the vortex, \eqref{gaugeCond} translates to the relation
 \begin{equation}
     \partial_i a_i = -q^2 \rho_s^2 l. \label{gaugeCondinf}
 \end{equation}{}
This identity now allows to completely eliminate $l$, making the system's three degrees of freedom manifest. This gauge also reduces the number of independent equations of motion: Using \eqref{gaugeCondinf}, it is easy to see that the equation of motion for $l$ can be obtained from the one for $a_i$ by acting with $\partial_i$ on both sides of the latter one. \newline
The next step of this discussion is to set up the description in terms of warp and twirl fields. It is easy to see from \eqref{LagrHiggs} that there exists a third non-trivial zero mode in the presence of the vortex, corresponding to a shift of $\lambda$ by an arbitrary constant, i.e. a shift of the scalar field's $U(1)$-phase. This zero mode can be promoted to a twirl field, thereby providing the missing third degree of freedom. This choice is particularly useful, as this field coincides with the field $l$, so that its full dynamics is captured in the warp field picture. \newline 
Following the arguments of the last subsection, the theory of the warp fields together with the twirl field can be described by a Lagrangian of the form \eqref{Righthere}. As the underlying theory is known, the missing coefficients can be determined by an explicit computation, yielding
\begin{equation}
\begin{split}
    C_{ab}^{\mu \nu} (\xi)=& \frac{1}{2} \eta^{\mu \nu} \big(\partial_a \rho_s \partial_b \rho_s + \frac{q^2 \rho_s^2 }{2}    \partial_a \lambda_s \partial_b \lambda_s  + \partial_a A_i^s \partial_b A_i^s \big) + \frac{1}{2} \partial_a A^\mu_s  \partial_b A^\nu_s , \\
    D_{a}^{\mu \nu} (\xi) =& \frac{q^2 \rho_s^2 }{2} \eta^{\mu \nu}, \ \ C_a^\mu(\xi)=  q^2 \rho_s^2  A_s^\mu \partial_a \lambda_s ,  \ \ D^\mu (\xi)=q^2 \rho_s^2  A_s^\mu ,\ \  g(\xi)= \frac{q^2 \rho_s^2 }{2}A_i^s  A_i^s,
    \label{CoeffHiggs}
\end{split}{}
\end{equation}{}
where all functions on the right hand side have to be understood as functions of $\xi$. \newline
Up to this point, this theory is conceptually very similar to the example discussed in section 3.3. The main difference lies now in the way it is related to the full theory. On the linearized level, the fluctuations induced by the warp and twirl fields are given by
\begin{equation}
\begin{split}
r^\varphi=& - \partial_a \rho_s \varphi^a, \\
l^\varphi=& - \partial_a \lambda_s \varphi^a + \vartheta, \ \text{and} \\
a^\varphi_i =& - \partial_a A^s_i \varphi^a. \label{RelHiggs} 
\end{split}{}
\end{equation}{}
Here, the label $\varphi$ implies that the quantities on the left hand side have to be understood as functions of $\{\varphi^a\}_a$ and $\vartheta$. These relations can now be used to show the equivalence of the two theories, in the sense that the equations of motions of the fluctuations can be used to obtain the ones of the warp and twirl fields and vice versa. \newline 
The first part of this proof is rather straightforward. Using \eqref{RelHiggs}, the equations of motion for the warp fields can be written as 
\begin{align}
    \rho_s^2 \ddot{l}^\varphi&= \partial_i \big(\rho_s^2 j^\varphi_i+2 \rho_s J_i^s r^\varphi \big)  \label{eomHiggstwirl} \\ 
    \partial_a \rho_s & \big(\Box r^\varphi +q^2 J_i^s J_i^s r + 2 q^2 J_i^s \rho_s j_i^\varphi + V^{\prime \prime} (\rho_s) r^\varphi \big)
    =\partial_a A_i^s \big( \partial_m f_{mi} - 2 q^2 J_i^s \rho_s r - q^2 \rho_s^2 j_i -  \ddot{a_i} \big) \label{eomHiggswarp}
\end{align}{}
Comparing these with \eqref{eomHiggs}, it is easy to see that these equations are a linear combination of the equations of motion of the full theory. Thus, a given configuration of fluctuations satisfying its equations of motion implies that the warp fields inducing said fluctuations are also solutions of the equations obtained from their theory. \newline 
The equations \eqref{eomHiggswarp} and \eqref{eomHiggstwirl} are also essential in the derivation of the equations of motion of the full theory from the ones of the warp and twirl fields. First of all, the equation for $\vartheta$ immediately yields the equation for $l$. Multiplying \eqref{eomHiggswarp} with the factor $\epsilon_{am}\hat{x}_m$ and making use of the explicit form of the vortex \eqref{Vortex}, one obtains the equation
\begin{equation}
    \hat{x}_i\big(  \ddot{a_i} - \partial_m f_{mi} + 2 q^2 J_i^s \rho_s r + q^2 \rho_s^2 j_i \big)=0.
\end{equation}{}
In two spatial dimensions, this implies that
\begin{equation}
     \ddot{a_i} - \partial_m f_{mi} + 2 q^2 J_i^s \rho_s r + q^2 \rho_s^2 j_i = \Gamma \epsilon_{im} \hat{x}_m, \label{Gammaeq}
\end{equation}{}
with some arbitrary function $\Gamma$. \newline 
A similar equation can be obtained for $r$. Multiplying both sides of \eqref{eomHiggswarp} $\hat{x}^a$ allows to handle the factor $\partial_a \rho_s$ on the left side of the equation. Further eliminating the $a_i$-terms on the right hand side using \eqref{Gammaeq}, one obtains the following equation for $r$:
\begin{equation}
    \Box r^\varphi +q^2 J_i^s J_i^s r + 2 q^2 J_i^s \rho_s j_i^\varphi + V^{\prime \prime} (\rho_s) r^\varphi= \Gamma \cdot \frac{1-f+r f^\prime}{q r^2 \nu \sigma^\prime (r)}.
\end{equation}{}
Up to the terms proportional to $\Gamma$, these last two equations coincide precisely with the equations of motion for $a_i$ and $r$ in the full theory. This new factor $\Gamma$ can now be eliminated using the gauge condition \eqref{gaugeCondinf}, which has been imposed on the field space of the fluctuations. Recall that \eqref{gaugeCondinf} implies that the divergence of the equation for $a_i$ is reduced to the equation for $l$. In a similar way, the left hand side of the equation for $a_i^\varphi$ is reduced to the left-hand side of the equation for $l^\varphi$, i.e. the equation of motion for $\vartheta$. As the latter one is independent of $\Gamma$, this leads to a constraint on $\Gamma$,
\begin{equation}
    \partial_i \Gamma \epsilon_{im} \hat{x}_m=0.
\end{equation}{}
In two spatial dimensions, this means that $\Gamma = \Gamma (r)$, which implies that $\Gamma$ cannot contain the fluctuations $a_i$, $r$ and $l$, as these are in general also functions of the angular coordinate. Thus, the right hand side of the equation for $r$ as well as for $a_i$ is of $0^{\text{th}}$ order in the fluctuation, which would correspond to a term linear in the fluctuations in the full theories underlying Lagrangian. As the fluctuations are perturbations around a solution of the equation of motion, such a term is forbidden by the equations of motion of the vortex. Thus, in this gauge, $\Gamma=0$, and the equations of motion of the full theory are reproduced. \newline
This implies now that if extended by a twirl field, the theory of warp fields is capable of capturing all degrees of freedom of the underlying theory. The construction of this theory itself is straightforward and its results are in perfect agreement with the arguments of the last subsection. This description further appears with the interesting feature that it allows for a redundance-free description of the system's dynamics. The main complication arising from this approach is encountered only if one tries to match the theory with the description in terms of the usual fluctuations. Its origin lies however not in the properties of the warp fields or the twirl fields respectively, but the underlying theory's remaining gauge redundancy.

\section{Conclusion \& Summary}

The concept of the moduli space allows for a compact description of the low-energy dynamics in the presence of a topological soliton. One of the central reasons for its importance lies thereby in its universality. It can be used to describe the low energy dynamics of any theory giving rise to a soliton, and all properties of the underlying theory as well as the soliton are fully encoded in the moduli space metric. \newline 
Warp and twirl fields can be understood as an attempt to extend this picture. These new fields can be used to describe the full dynamics of any theory - including the sector corresponding to its non-zero modes - giving rise to a soliton nearby the latter one, and all properties of the underlying theory as well as the soliton are fully encoded in a small number of background functions. This includes the precise shape of the soliton, the types of fields involved in its formation etc. Herein lies the power of this approach, as it not only reveals a structure hidden in a wide range of models for which it provides a compact description, but also allows for an alternative interpretation of a significant subsector of these theories. \newline 
One aspect of this new perspective is the geometric description presented in section 2.7 as well as the duality transformation that it is based on. While it has a quite simple interpretation, it also leads to the in this context rather unexpected appearance of the Einstein-Hilbert action as well as a field sourced by the soliton's energy-momentum tensor. \newline 
The main constructions presented in this article are particularly interesting from the perspective of effect field theory. By performing a simple modification of the low-energy theory, extending the domain of some of its degrees of freedom, one obtains a theory applicable at all energies covered by the underlying theory. The condition of low energy becomes fully replaced by a constraint on the localization of the configurations of interest.

\section*{Acknowledgements}
This work was made possible by a scholarship of Heinrich-Böll-Stiftung, which I thank for years of formative support. I further thank my supervisor Stefan J. Hofmann for advice and an inspiring working environment, as well as Michael Haack, Emmanouil Koutsangelas, Marc Schneider and Maximilian Koegler for proofreading several early versions of this article. I especially thank Gia Dvali for particularly enriching discussions as well as for helpful remarks and literature.
	
\appendix

\section{Appendix: Quantum theory of the warp field}
	
	The quantum theory of $\varphi$ is determined by the properties of the operators $\hat{\varphi}$, $\hat{\varpi}$ and the Fock space $\mathcal{F}$ they are acting on. Just as in the classical case, it can be embedded into the $\phi$-theory.  \newline
	The following discussion is limited to the free theory. The main motivation therefore lies again in the technical difficulties of the non-linear theory, which are amplified due to the unusual structure of the Fock space. 
	
	\subsection{Linearized theory}
	The operators representing $\varphi$ and $\varpi$ can be obtained from \eqref{Solvp}:
	\begin{equation}
	\begin{split}
	\hat{\varphi} (t,x)= &  \mu_m \hat{v} (t -t_0 ) + \mu_m \hat{\zeta} + \sumint_k  \hat{\alpha}_k^* e^{i \omega_k t} g_k^* (x) + \hat{\alpha}_k e^{-i \omega_k t} g_k (x) \\
	\hat{\varpi} (t,x)= & i \nu^2 (\sigma^\prime)^2 (\mu_m x) \cdot \bigg( \mu_m \hat{v} + \sumint_k  \hat{\alpha}_k^\dagger \omega_k e^{i \omega_k t} g_k^* (x) - \hat{\alpha}_k \omega_k e^{-i \omega_k t} g_k (x)  \bigg) 
	\label{Opvp}
	\end{split}{}
	\end{equation}{}
	These operators satisfy the usual equal time commutation relations,
	\begin{equation}
	\begin{split}
	[\hat{\varphi} (t,x),\hat{\varpi} (t,x^\prime)]=&i \delta (x-x^\prime) \ \text{and}  \\
	[\hat{\varphi} (t,x),\hat{\varphi} (t,x^\prime)]=&0=[\hat{\varpi} (t,x),\hat{\varpi} (t,x^\prime)].
	\end{split}{}
	\end{equation}{}
	In terms of the creation/annihilation operators $\{\alpha_k^{(\dagger)}\}_k$ and $\hat{v}$, $\hat{\zeta}$, these relations can be expressed as
	\begin{equation}
	\begin{split}
	[\hat{\alpha}_k , \hat{\alpha}_p^\dagger] =& \begin{cases}
	\mathcal{N}_c  \delta (k-p), & \text{for } k,p \in \mathcal{I}_c \\
	\mathcal{N}_d \delta_{kp}, & \text{for } k,p \in \mathcal{I}_d
	\end{cases}, \\
	[\hat{\zeta}, \hat{v}]=& i M_{\text{sol}}^{-1},
	\label{Comm}
	\end{split}{}
	\end{equation}{}
	where $\mathcal{I}_c$ and $\mathcal{I}_d$ denote the continuous and discrete parts of the spectrum. The normalization constants $\mathcal{N}_c$ and $\mathcal{N}_d$ are determined by the chosen normalization of the mode functions and eigenstates. Defining $\hat{p}=M_{\text{sol}} \hat{v}$, the last relation is equivalent to the commutation relation satisfied by the collective coordinate and its conjugate momentum. A similar treatment is possible in the $\phi$-theory, as disussed in \cite{creutz1975}. \newline
	The Fock space these operators act on is therefore given by
	\begin{equation}
	\begin{split}
	\mathcal{F}=& L^2 (\mathbb{R},\text{d}x) \otimes \mathcal{F}_p, \ \text{where} \\ 
	\mathcal{F}_p=&\bigoplus_{n=0}^{\infty} S_n \big(L^2 (\mathbb{R},\text{d} \varsigma)\big)^{\otimes n}
	\end{split}
	\end{equation}
	denotes the Fock space of particle-like excitations and $L^2 (\mathbb{R},\text{d}x)$ corresponds to the collective coordinate. Following the usual naming scheme of particle physics, the particle-like excitations represented by this Fock space will be named \textit{warpions}. \newline
	The chosen Fock space differs from one of a usual scalar field due to its spatial measure d$\varsigma$, which is an immediate result of the mode functions $\{g_k (x)\}_k$. To see this, consider the Fock space creation/annihilation operators $\mathcal{A}^{(\dagger)}$, which are related to the $\alpha^{(\dagger)}$-operators via
	\begin{align}
	\mathcal{A}^{(\dagger)}[f]= \sumint_k f^k \alpha^{(\dagger)}_k,
	\end{align}{}
	where $f^k$ denotes the coefficients of $f$ when expanded in the $\{g_k\}_k$-basis. These operators satisfy the commutation relation
	\begin{align}
	[\mathcal{A}[f], \mathcal{A}^\dagger[g]] = (f,g)_{\mathcal{H}}, \label{Comm2}
	\end{align}{}
	where $(f,g)_{\mathcal{H}}$ denotes the scalar product on the one particle Hilbert space $\mathcal{H}$ \cite{Fin}. As $\alpha_k^{(\dagger)} = A^{(\dagger)} [g_k]$, \eqref{Comm} and \eqref{Comm2} cannot be simultaneously realized on the usual $(L^2,\text{d}x)$. But as the mode functions $\{g_k \}_k$ are orthonormal with respect to the measure $\text{d} \varsigma$, the desired commutation relations are realized on $\mathcal{H} = (L^2 ,\text{d} \varsigma )$. \newline
	These relations can now be used to obtain an expression for the propagator of the warp field. The non-zero mode contributions can be obtained from the expectation value of the time-ordered product of the non-zero mode terms within $\hat{\varphi}$ with respect to the $\mathcal{F}_p$-vacuum. In terms of the $\phi$-propagator, they can be expressed as 
	\begin{align}
	\Delta_{\varphi}^{nz} \big( (t,x) , (t^\prime ,x^\prime) \big)= \frac{ \Delta_{\phi} \big( (t,x) , (t^\prime ,x^\prime) \big)}{\big(\nu \sigma^\prime (\mu_m x)\big) \big(\nu \sigma^\prime (\mu_m x^\prime)\big)}. \label{Propagator}
	\end{align}{}
	The propagator of the zero mode is in great detail discussed in \cite{creutz1975}. 
	
	\subsection{Implications of the probability measure}
	Similar to the classical theory, the quantum theory of the warp field bears a strong resemblance to the theory of a usual scalar field, with the main differences being the appearance of the weight function $\rho (x)= \nu^2 (\sigma^\prime)^2 (\mu_m x)$ and the unusual asymptotic behavior of the mode functions. The quantum theory however reveals the true relevance of these functions, namely that they enforce a probability measure which is proportional to the weight function $\rho$. This is nothing but a particle physics version of the localization behavior observed in the classical theory. While warpion-wave functions are supported on all of spacetime, the associated probability density is localized in the same region as the soliton as it is weighted with $\rho$. \newline
	Due to the incorporation of $\rho$ in the measure of $\mathcal{H}$ this localization is also carried over to the expectation values of any observable. Let $\mathcal{O}$ denote some observable which for simplicity acts trivially on the $L^2$-space of the zero mode. In terms of the corresponding one-particle Hilbert space observable $\hat{O}$, the action of this operator on some Fock space state $\ket{\Psi}= \ket{\Psi_1 (x), \Psi_2 (x_1,x_2), \dots}$ is given by
	\begin{equation}{}
	\begin{split}{}    
	\mathcal{O} \ket{\Psi}&= \ket{ (\mathcal{O}\Psi)_1 (x), (\mathcal{O}\Psi)_2 (x_1,x_2), \dots}, \ \text{where} \\
	\big(\mathcal{O} \Psi\big)_n & (x_1, \dots x_n)= \sum_{i=1}^{n} \hat{O}(x_i) \Psi_n (x_1, \dots x_n).
	\end{split}{}
	\end{equation}{}
	The functional dependence of $\hat{O}(x_i)$ is here meant as the action of $\hat{O}$ on the $i^{\text{th}}$ argument of $\Psi_n$. Therefore, the expectation value takes the form
	\begin{equation}{}
	\begin{split}{} 
	\bra{\Psi}\mathcal{O}\ket{\Psi}= & \braket{\Psi_1|(\mathcal{O}\Psi)_1} + \braket{\Psi_2|(\mathcal{O}\Psi)_2}+ \dots= \\ 
	=& \intd \varsigma (x) \ \Psi_1^* (x) \hat{O} \Psi_1 (x) + \\ 
	&+\intd \varsigma (x_1) \intd \varsigma (x_2) \ \sum_{i=1}^{2} \Psi_2^* (x_1, x_2) \hat{O}(x_i) \Psi_2 (x_1, x_2) + \dots
	\end{split}{}
	\end{equation}{}
	The important feature of this expression is now that the contribution of each $n$-excitation state comes with $n$ integrals, each corresponding to one of the excitations. As the measure of integration is given by d$\varsigma$ and therefore contains the function $\rho$, the contribution of each warpion to any physical observable is again weighted by the density of the soliton.

	\subsection{Embedding into the full linearized theory}
	Just as in the classical case, the zero mode of the warp field can immediately be identified with the collective coordinate as introduced in \cite{creutz1975}. To extend this insight to the non-zero modes, consider the following map between the one particle Hilbert spaces of the $\varphi$ and the $\phi$ theory:
	\begin{equation} 
	\begin{split}
	\iota : \mathcal{H} & \rightarrow \mathbb{H} \\ 
	\Psi (x) & \mapsto \rho^{1/2} (x) \Psi (x) \label{iota}
	\end{split}
	\end{equation}
	This map is an isometry which respects the structure of the mode expansion, in the sense that $\iota (g_k) = f_k$ and $a_k = \iota \alpha_k \iota^{-1}$. \newline
	Extending this map to the whole Fock space in the canonical way \cite{Fin}, it is clear that the dynamics on $\mathcal{F}_p$ generated by the $\{ \alpha_k^\dagger \}_k$ operators is equivalent to the one on the corresponding subset of $\mathbb{F}$ generated by the $\{ a_k^\dagger \}_k$ operators.


\begin{thebibliography}{9}
		
		\bibitem{Proton}
		G. S. Adkins, C. R. Nappi, E. Witten, \textit{Static Properties of Nucleons in the Skyrme Model}, \textit{Nuclear Physics B} \textbf{228} (1983), 552
		
		\bibitem{Skyrme}
		J. Schechter, H. Weigel, \textit{The Skyrme Model for Baryons}. (1999) [arXiv:hep-ph/9907554v1]
		
		
		
		\bibitem{Casimir}
		B. Moussallam, \textit{Casimir Energy in the Skyrme Model}. (1992) [arXiv:hep-ph/9211229v1]
		
		\bibitem{Dvali}
		G. Dvali, M. Shifman, \textit{Dynamical Compactification as a Mechanism of Spontaneous Supersymmetry Breaking}, \textit{Nuclear Physics B} \textbf{504} (1997) 127, [hep-th/9611213.7]
		
		\bibitem{Arkani}
		N. Arkani-Hamed, S. Dimopoulos, G. Dvali, \textit{The Hierarchy Problem and New Dimensions at a Millimeter}, \textit{Physics Letters B} \textbf{429} (1998), 263 [hep-ph/9803315.8]
		
		\bibitem{Randall}
		L. Randall, R. Sundrum, \textit{An Alternative to Compactification}, \textit{Physics Review Letters} \textbf{83} (1999), 4690 [hep-th/9906064]
		
		\bibitem{Arkani2}
		N. Arkani-Hamed, S. Dimopoulos, G. Dvali, \textit{Phenomenology, Astro-physics and Cosmology of Theories with Sub-Millimeter Dimensions and TeVScale Quantum Gravity}, \textit{Physics Review D} \textbf{59} (1999), 096004 [hep-ph/9807344]
		\bibitem{Rubakov}
		V.A. Rubakov, M.E. Shaposhnikov, \textit{Do We Live Inside A Domain Wall?}, \textit{Physics Letters B} \textbf{125} (1983), 1366
		
		\bibitem{Brane}
		J. Polchinski, \textit{Lectures on D-Branes} (1997) [hep-th/9611050v2]
		
		\bibitem{creutz1975}
		M. Creutz, \textit{Quantum mechanics of extended objects in relativistic field theory}, \textit{Physical Review D} \textbf{12} (1975), 3126
		\bibitem{BaackeRothe}
		J. Baacke, H. J. Rothe, \textit{On the Quantization of Moving Extended Objects in One Space Dimension}, \textit{Nuclear Physics B} \textbf{118} (1977), 371
		\bibitem{goldstone1975}
		J. Goldstone, R. Jackiw, \textit{Quantization of nonlinear waves},
		\textit{Physical Review D} \textbf{11} (1975), 1486
		\bibitem{Faddeev1976}
		L.D. Faddeev, V.E. Korepin, \textit{About the zero mode problem in the quantization of solitons}, \textit{Physics Letters} \textbf{63B} (1976), 435
		\bibitem{jevicki1976}
		A. Jevicki, \textit{Treatment of zero-frequency modes in perturbation
			expansion about classical field configuratios}, \textit{Nuclear Physics B} \textbf{117} (1976), 365
		
		
		\bibitem{jackiw1977}
		R. Jackiw, \textit{Quantum meaning of classical field theory}, \textit{Review of Modern Physics} \textbf{49} (1977), 681
		
		
		
		
		\bibitem{Rajaraman} 
		R. Rajaraman. 
		\textit{Solitons and Instantons: An Introduction to Solitons and Instantons in Quantum Field Theory}. 
		North-Holland (1982).
		\bibitem{Shifman} 
		M. Shifman. 
		\textit{Advanced topics in Quantum Field Theory}. 
		Cambridge University Press (2012).
		\bibitem{Manton}
		N. Manton, P. Sutcliffe.
		\textit{Topological Solitons}.
		Cambridge monographs on mathematical physics (2004).
		
		
		
		
		\bibitem{PfadInt1}
		J.L. Gervais, A. Jevicki, B. Sakita.
		\textit{Phys. Rev. D} \textbf{12}, 1038 (1975)
		
		\bibitem{PfadInt2} J.L. Gervais, A. Jevicki, B. Sakita,
		\textit{Phys. Rep} \textbf{23}, 281 (1976)
		\bibitem{PfadInt3}
		R. F. Dashen, E. Hasslacher, A. Neveu.
		\textit{Physical Review D} \textbf{10}, 4114 (1974)
		
		
		
		
		
		\bibitem{goldstone1962}
		J. Goldstone, A. Salam, S. Weinberg, \textit{Broken Symmetries}, \textit{Physical Review} \textbf{127} (1962), 965
		
		\bibitem{Sred}
		M. Srednicki.
		\textit{Quantum Field Theory}.
		Cambridge University Press (2007).
		
		\bibitem{Peskin}
		M. Peskin, D. Schroeder.
		\textit{An Introduction To Quantum Field Theory}.
		CRC Press (1995).
		
		\bibitem{DvaliShifman}
		G. Dvali, M. Shifman, \textit{Tilting the Brane, or Some Cosmological Consequences of the Brane Universe}, \textit{Phys.Rept.} \textbf{320}, (1999) [hep-th/9904021v1]
		
		
		
		
		\bibitem{Stefan}
		S. Hofmann, M. Schneider, \textit{Classical versus quantum completeness}, \textit{Phys. Rev. D} \textbf{91} (2015), 125028 [hep-th/1504.05580v2]
		
		
		
		\bibitem{Fin}
		G. Scharf
		Finite Quantum Electrodynamics: The Causal Approach, Third Edition
		Dover Books on Physics (2014)
		
		
		
		
		\bibitem{Weinberg} 
		S. Weinberg. 
		\textit{Gravitation and Cosmology: Principles and applications of the general theory of relativity}. 
		Wiley (1972).
		
		\bibitem{Skyrme0}
		T. H. R. Skyrme, \textit{A Non-Linear Field Theory}, \textit{Proc. Roy. Soc.} \textbf{127} (1961) 
		
		\bibitem{Nielsen}
		H. B. Nielsen, P. Olesen, \textit{Vortex-line models for dual strings}. \textit{Nuclear Physics B}, \textbf{61} (1973) 
		
		\bibitem{constrHam}
		A. J. Hanson, T. Regge, C. Teitelboim. \textit{Constrained Hamiltonian Systems}. Academia Nazionale Dei Lincei (1976)
	\end{thebibliography}
\end{document}